**The Effectiveness of Virtual Patient Simulation Versus Peer Simulation in Providing Sexual Counseling During Pregnancy: A Randomized Controlled Trial**

Neslihan YILMAZ SEZER[a], Menekşe Nazlı AKER[b], Pinar KULLU[c], Rana Tuncer[d], Gael Lucero-Palacios[e], Roghayeh Leila BARMAKI[f]

[a] PhD, Ankara University, Faculty of Nursing, Midwifery Department, Ankara, Turkey nyilmaz@ankara.edu.tr

[b] PhD, Ankara University, Faculty of Nursing, Midwifery Department, Ankara, Turkey menekseaker@gmail.com

[c] PhD, University of Delaware, Department of Computer and Information Sciences, Newark, DE, United States pinarkullu@gmail.com

[d] BSc, University of Delaware, Department of Computer and Information Sciences, Newark, DE, United States rantun@udel.edu

[e] BSc, University of Delaware, Department of Computer and Information Sciences, Newark, DE, United States gael@udel.edu

[f] PhD, University of Delaware, Department of Computer and Information Sciences, Newark, DE, United States rlb@udel.edu

**Correspondence**

Menekşe Nazlı Aker, Assistant Professor, Ankara University Faculty of Nursing, Ankara, Turkey

To whom correspondence should be addressed at Hacettepe Mahallesi Plevne Caddesi No:7 PK: 06230 Altındağ / ANKARA, TURKEY; e-mail: menekseaker@gmail.com; Telephone: +9(0) 0312 319 1450/ 2831

**(1) Conflict of Interest**

We wish to confirm that there are no known conflicts of interest associated with this publication.

**(2) Ethical Approval**

The study was approved by the Ethics Committee of Ankara University (Tarih:03.03.2025 Sayı:05/87).

**(3) Funding Sources**

This research was supported by Ankara University Scientific Research Projects Coordination Unit with project ID 4006. Additional support was provided by the U.S. National Science Foundation (NSF) through the Collaborative Research: FW-HTF-P project (Award Nos. 2222661 and 2222663).

All listed authors meet the authorship criteria and all authors are in agreement with the content of the manuscript.

**(4) Data Availability Statement**

The data that support the findings of this study are not publicly available due to ethical and privacy considerations. However, anonymized data may be made available from the corresponding author upon reasonable request and subject to appropriate ethical approval.

**Acknowledgements**

Ankara University Scientific Research Projects Coordination Unit is acknowledged for funding. The U.S. National Science Foundation Research Experiences for Undergraduates (NSF REU) program is also acknowledged for its support.

**ABSTRACT**

**Background:** Although sexual health counseling is one of the important responsibilities of healthcare professionals, effective educational methods are needed to develop students' counseling skills in this field. The aim of this study was to compare the effectiveness of virtual patient simulation and peer simulation methods in developing sexual counseling skills during pregnancy among nursing faculty students.

**Methods:** This randomized controlled trial included 51 participants assigned to one of three groups: virtual patient simulation (n = 17), peer simulation in a virtual environment (n = 17), or face-to-face peer simulation (n = 17). In the study, all groups received face-to-face theoretical instruction on sexual counseling during pregnancy. Following the theoretical training, students participated in virtual patient simulation, peer simulation in a virtual environment, or face-to-face peer simulation practices according to the groups to which they were assigned by randomization. Outcome measures included the Sexual Attitudes and Beliefs Scale (SABS), the Student Satisfaction and Self-Confidence in Learning Scale, and the Sexual Counseling Skills Evaluation Form.

**Results:** It was determined that the participants' SABS scores decreased significantly after the intervention. According to the results of the mixed repeated-measures ANOVA, the effect of time was statistically significant ($F = 113.31$, $p < 0.001$). However, the group effect ($F = 0.836$, $p = 0.440$) and the group × time interaction ($F = 0.304$, $p = 0.739$) were not significant. There was no difference between the groups in terms of Satisfaction in learning, Self-confidence in learning, and Skill scores ($p > 0.05$).

**Conclusion:** This study showed that different simulation methods used in sexual counseling education during pregnancy were effective in reducing students' negative attitudes and beliefs and provided similar educational outcomes. Therefore, virtual patient and peer simulations may be recommended as feasible approaches for improving sexual counseling skills.

## 1. INTRODUCTION

The provision of education and counseling services to patients by healthcare professionals is considered a professional, ethical, and legal requirement (Bensing et al., 2001; Zolkefli & Mahmud, 2023). In particular, midwives and nurses play a key role in patient education and counseling services because they are in continuous and direct communication with patients (Açıkgöz & Baykal, 2023; Altınayak et al., 2020; González et al., 2005; Pouresmail et al., 2023). However, in order for these responsibilities to be carried out effectively, healthcare professionals need to be equipped with adequate knowledge, skills, and attitudes. Developing counseling and communication skills before graduation supports the readiness for practice, self-efficacy, and professional competence of healthcare professional candidates (Kienle et al., 2021; Shatto et al., 2022; Yang & Kim, 2022). In this process, it is particularly important to focus on sensitive topics such as sexuality and sexual health. Indeed, studies show that healthcare professionals have difficulty discussing these topics and are reluctant to gain sufficient practical experience (Fennell & Grant, 2019; Kelder et al., 2022; Kong et al., 2009). Lack of knowledge and skills is shown among the main reasons why healthcare professionals have difficulty providing counseling in the field of sexual health (Bogaert & Roels, 2025; Fennell & Grant, 2019; Manninen et al., 2024; Verrastro et al., 2020). Insufficient inclusion of sexual health topics in the education of healthcare professional candidates and inadequate development of counseling skills may lead to deficiencies in responding to individuals' sexual health-related needs after graduation, which may negatively affect the quality of sexual health services (Ferrara et al., 2003; Manninen et al., 2022; Merhavy et al., 2023). Therefore, it is important to develop counseling skills at an early stage. In the literature, studies have shown that standardized patients, role-playing, peer simulation, and technology-based approaches such as avatars and virtual patients are used to improve students' counseling skills (Bracq et al., 2019; Lane & Rollnick, 2007; Magill et al., 2022; Sezer et al., 2023).

Peer simulation is an educational method that enables students to learn from one another, promotes collaboration, and improves communication skills. Frequently preferred in practical applications, this method allows students both to share knowledge and to experience their mistakes in a safe environment (Henning et al., 2008; Yoo & Chae, 2011). However, peer simulation requires an intensive preparatory process. In addition, differences in students' learning pace and levels of knowledge may complicate the learning process (Şenyuva & Akince, 2020). Therefore, innovative approaches that may serve as alternatives to peer simulation are increasingly being adopted.

Virtual patient simulations are computer-based programs in which clinical scenarios are designed to replicate real clinical encounters. They provide users with experience in virtual environments in areas such as encountering patients in virtual settings, taking patient histories, performing physical examinations, making treatment decisions, and providing education and counseling, thereby facilitating the educational process (Yılancıoğlu & Bildik, 2022). The advantages of virtual patient simulations include repeatability, a safe learning environment, and the opportunity for students to encounter different scenarios. Particularly in sensitive topics such as sexual health, virtual environments enable students to improve their counseling skills while practicing without fear of making mistakes (Kononowicz et al., 2019; Peddle et al., 2016). This study aimed to compare the effectiveness of virtual patient simulation and peer simulation in developing sexual counseling skills during pregnancy among nursing faculty students. The topic was selected because pregnancy is one of the periods in which sexual counseling services are most needed. Physiological changes, hormonal fluctuations, and misconceptions about sexuality experienced during this period may negatively affect couples' sexual lives; therefore, sexual health counseling contributes to couples maintaining a healthy sexual life and strengthening their relationship dynamics (Fathalian et al., 2022; Shahbazi et al., 2019; Ziaei et al., 2022).

## 2. METHOD

### Aim of the Study

The aim of this study was to compare the effectiveness of virtual patient simulation and peer simulation methods in developing sexual counseling skills during pregnancy among nursing faculty students.

### Type of Study

This study was a single-center, parallel-group, randomized controlled experimental study. Figure 1 shows the CONSORT (Consolidated Standards of Reporting Trials) diagram. This study was registered in the Clinical Trials registry system with the registration number NCT06949501.

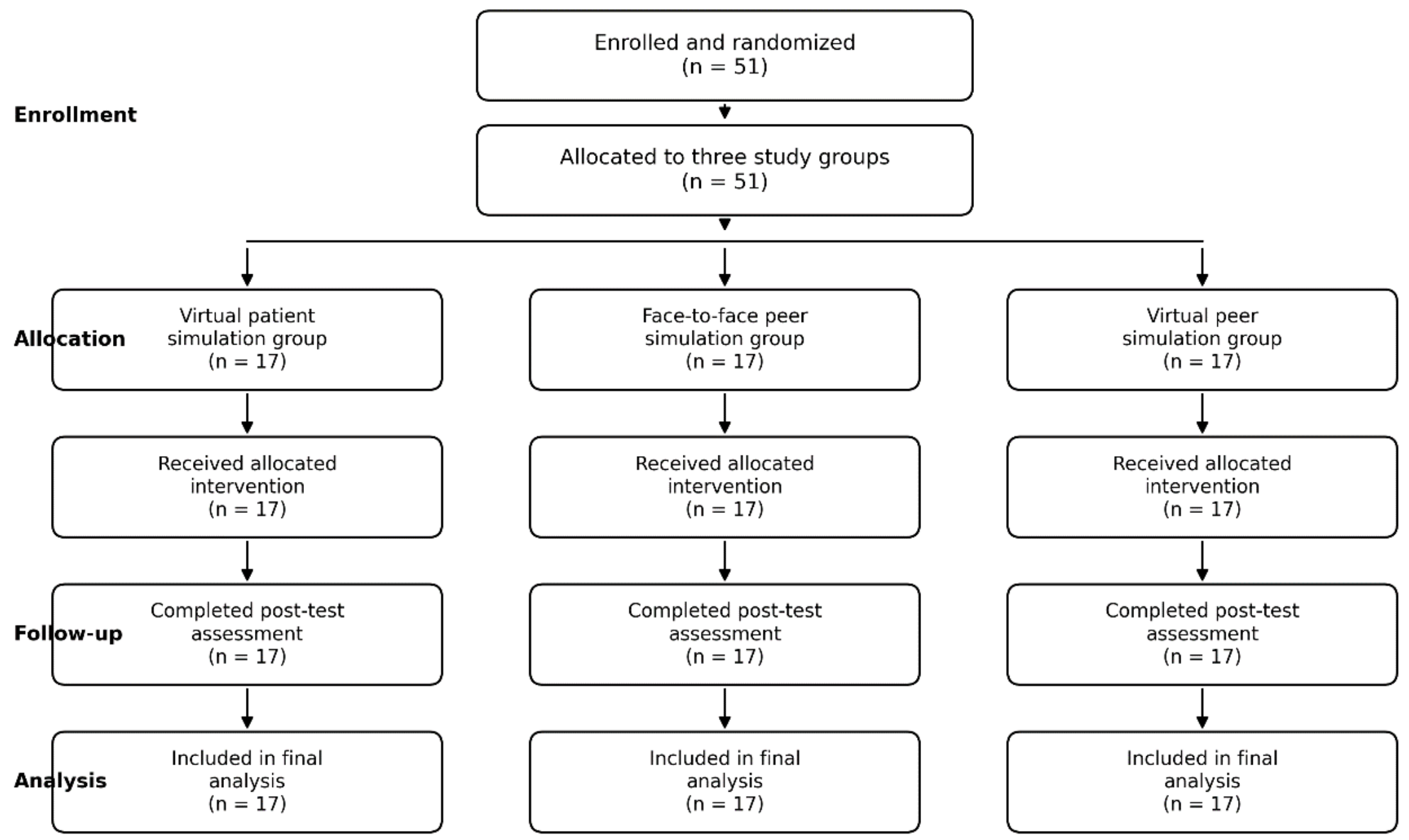

**Figure 1. CONSORT Flow Diagram**

**Participants**

This study was conducted between April and June 2025 with students enrolled in the Faculty of Nursing at a university. A power analysis was performed to determine the sample size of the study. The power analysis was conducted using the G*Power 3.1.9.7 program based on one-way analysis of variance (ANOVA: fixed effects, omnibus, one-way). Based on the results of the study by Tugut and Golbasi (2017), the effect size was accepted as f = 0.74 (Tugut & Golbasi, 2017). As a result of the calculation performed with an alpha error level of α = 0.05, 99% power, and three groups, the minimum sample size required for the study was determined as a total of 45 students. Considering possible participant attrition during the research process, 10% was added to the calculated sample size. Accordingly, 17 students were included in each group, and the study was completed with a total of 51 students (Figure 1). Students who had completed the required clinical coursework related to maternal health care and agreed to participate in the study were included. In this study, students who had successfully completed the Prenatal Period and Practice in Normal Birth course or the Gynecology and Obstetrics Nursing course were considered to have the ability to provide care to pregnant women. Students who had frozen their registration or suspended their education, as well as those who had previously received formal training on providing sexual counseling, were not included in the study.

**Randomization**

Students who met the inclusion criteria were assigned to three groups: the virtual patient simulation group, the peer simulation group in a virtual environment, and the face-to-face peer simulation group. Simple randomization was used to ensure similarity between the groups and to reduce selection bias. Sealed envelopes containing group assignments were prepared to ensure that the randomized students were assigned to the groups in an unbiased manner. Lots were drawn by an independent researcher who was not involved in the study, and the groups in which the students would take part were determined. Outcome assessor blinding was performed to prevent detection bias.

**Parameters Examined in the Study**

In the study, students' descriptive characteristics, sexual attitudes and beliefs, sexual counseling skills, satisfaction in learning, and self-confidence in learning were evaluated.

***Descriptive Characteristics of the Students***

The descriptive characteristics of the students were evaluated using the Student Descriptive Information Form. This form consisted of questions regarding students' age, economic status, academic grade point average, and previous experience in providing sexual counseling. The form was administered to the students at the beginning of the study.

***Sexual Attitudes and Beliefs***

Students' sexual attitudes and beliefs were evaluated twice using the Sexual Attitudes and Beliefs Scale (SABS): after group allocation and following completion of the simulation activities. The scale was developed by Reynolds and Magnan in 2005 (Reynolds & Magnan, 2005), and its Turkish validity and reliability study was conducted by Ayhan et al. in 2010. The scale consists of 12 items. The lowest possible score on the six-point Likert-type scale is 12, and the highest possible score is 72. A higher score on the scale indicates increased negative attitudes and beliefs regarding sexual care. The Cronbach's alpha coefficient of the scale was reported as 0.73 (Ayhan et al., 2010).

***Satisfaction and Self-Confidence in Learning***

Students' satisfaction in learning and self-confidence in learning were evaluated once, immediately after the simulation practices were completed, using the Student Satisfaction and Self-Confidence in Learning Scale. The scale was developed by Jeffries and Rizzolo in 2006 (Jeffries & Rizzolo, 2006). Its Turkish adaptation, validity, and reliability study was conducted by Unver et al. (2017). The scale consists of two subdimensions: satisfaction in learning and self-confidence in learning. It is a five-point Likert-type scale. As the score obtained from the scale increases, students' levels of satisfaction in learning and self-confidence in learning increase. The Cronbach's alpha coefficient of the scale was reported as 0.85 for the satisfaction in learning subdimension and 0.77 for the self-confidence in learning subdimension (Unver et al., 2017).

### *Sexual Counseling Skills*

Students' sexual counseling skills were evaluated by the same researcher using the Sexual Counseling Skills Evaluation Form while the students demonstrated their counseling skills. The form was prepared based on the BETTER, PLISSIT, Ex-PLISSIT, and One-to-One models (Annon, 1976; Borms & Vermeire, 2020; Bozkurt Karalı & Özkan, 2020; Mick et al., 2004; Taylor & Davis, 2006). The form, which includes all stages of the counseling process from initiating to terminating the interview, consists of 32 items. Each item was evaluated by assigning 0 or 1 point according to whether the student demonstrated the relevant behavior. The lowest possible score on the form is 0, and the highest possible score is 32. A higher score indicates a higher level of sexual counseling skill in the student.

## Implementation of the Study

In the study, the topic of sexual counseling during pregnancy was delivered theoretically and face-to-face to all groups. Following the theoretical training, students participated in virtual patient simulation, peer simulation in a virtual environment, or face-to-face peer simulation practices according to the groups to which they were assigned by randomization.

### *Theoretical Training*

All students included in the study received a standardized theoretical training. The theoretical training was conducted in four sessions, each lasting 50 minutes. The training content included knowledge of sexuality, sexuality during pregnancy, the counseling process, communication techniques, and sexual counseling models necessary for providing sexual counseling to pregnant women (Table 1).

All students participating in the study received the theoretical training at the same time. The training was conducted using a didactic lecture method and was supported by PowerPoint presentations, relevant visuals, and models.

**Table 1. Theoretical Training Program**

| Sessions | Training content |
|---|---|
| Session 1 | Genital anatomy; concepts related to sexuality |
| Session 2 | Physiology of sexual activity during pregnancy; factors affecting sexual activity during pregnancy; alternative ways other than sexual intercourse; sexual intercourse positions during pregnancy; misconceptions about sexuality during pregnancy; conditions in which sexual intercourse is restricted during pregnancy; importance of open communication between couples |
| Session 3 | Counseling process; essential characteristics required in counseling; verbal and nonverbal communication techniques |
| Session 4 | Use of models in counseling; case examples in sexual counseling |

**Practice Phase**

All students were included in the practice in the same communication skills laboratory, which was designed similarly to an outpatient clinic room. In the study, four different scenarios were developed for the practice of sexual counseling during pregnancy:

- A pregnant woman at the 16th week of pregnancy who avoids sexual intimacy because she is concerned that sexuality may harm herself and her baby,
- A pregnant woman at the 16th week of pregnancy who states that her partner avoids sexuality due to fear of harming the baby or the pregnant woman,
- A pregnant woman at the 32nd week of pregnancy who experiences sexual reluctance due to factors such as breast tenderness, pelvic congestion, and positional difficulties caused by the growing uterus,
- A pregnant woman at the 32nd week of pregnancy who avoids sexuality because she does not find herself attractive and thinks that her partner will not desire her.

After the students were randomized into groups, they were assigned a scenario by drawing lots. All students were expected to initiate the interview in accordance with the scenario assigned to them, provide sexual counseling, and appropriately terminate the interview at the end of the simulation.

In all simulation groups, the same researcher observed the student from a glass observation booth that allowed the simulation room to be monitored and completed the Sexual Counseling Skills Evaluation Form during this process. After the simulation was completed, feedback was provided to the students by the instructor, and the students completed the post-test forms after the practice.

***Virtual Patient Simulation Group***

Students in the virtual patient simulation group provided sexual counseling during pregnancy to the virtual patient model following the theoretical training. The virtual patient model was structured based on four scenarios. In line with the predetermined scenarios, the model was able to ask questions to the students, provide synchronous responses and verbal reactions to the students' questions, and enable synchronous communication. The virtual patient, positioned in the outpatient clinic room, was used to provide students with a realistic counseling experience. Students logged into the virtual patient simulation system using the password assigned to them.

***Peer Simulation Group in a Virtual Environment***

Students in the peer simulation group in a virtual environment provided sexual counseling during pregnancy to one of their peers in a virtual environment. In the peer simulation conducted in the virtual environment, students viewed the same background used in the virtual patient simulation; however, instead of interacting with the virtual patient, they connected with a peer through an online platform and carried out the counseling practice. The peer was trained in accordance with the four scenarios developed by the researchers and also used in the virtual patient simulation, and participated in a pilot practice together with the researchers. All peer roles were portrayed by the same trained student.

***Face-to-Face Peer Simulation Group***

Students in the face-to-face peer simulation group provided sexual counseling during pregnancy to a peer in a face-to-face setting. The peer was trained in accordance with the four scenarios developed by the researchers and also used in the virtual patient simulation, and participated in a pilot practice together with the researchers. All peer roles were performed by the same student who also took part in the peer simulation in the virtual environment.

**Data Evaluation**

The data obtained from the study were analyzed using the Statistical Package for the Social Sciences (SPSS) version 30.0 software package. Descriptive statistics and parametric and nonparametric tests appropriate to the characteristics of the variables were used in the evaluation of the data. The normality of continuous variables was assessed by examining skewness and kurtosis values, and values within the range of ±2 were considered indicative of normal distribution. In descriptive statistics, continuous variables were presented as mean ± standard deviation and median (minimum–maximum), while categorical variables were presented as number and percentage. In the comparison of the groups in terms of descriptive characteristics, one-way analysis of variance (One-Way ANOVA) was used for continuous variables, and the chi-square test was used for categorical variables. In comparisons among the three groups, One-Way ANOVA was applied for normally distributed continuous variables. In comparisons among the three groups, the Kruskal–Wallis test was applied for continuous variables that did not show normal distribution. Mixed repeated-measures ANOVA was used to evaluate the change in SABS scores over time and the group–time interaction. The level of statistical significance was accepted as $p < 0.05$.

**Ethical Considerations of the Study**

The study was approved by the Ethics Committee of Ankara University (Date: 03.03.2025; No: 05/87). In addition, written permissions were obtained from the faculty where the study was conducted. Written informed consent was obtained from all participants. The research was conducted in accordance with the Declaration of Helsinki.

**3. RESULTS**

Table 2 compares the descriptive characteristics of the groups. No statistically significant difference was found between the groups in terms of age, academic grade point average, economic status, or previous experience in providing counseling ($p > 0.05$).

**Table 2. Comparison of the Groups in Terms of Descriptive Characteristics**

| Variables | Virtual patient group (n = 17) | Face-to-face peer group (n = 17) | Virtual peer group (n = 17) | Test value / p |
|---|---|---|---|---|
| Age, Mean ± SD | 22.18 ± 0.72 | 22.06 ± 0.82 | 22.41 ± 0.93 | F = 0.786; p = 0.461 |
| Academic grade point average, Mean ± SD | 3.31 ± 0.18 | 3.36 ± 0.16 | 3.30 ± 0.19 | F = 0.443; p = 0.644 |
| Economic status, n (%) | | | | $\chi^2$ = 4.163; p = 0.125 |
| Moderate | 17 (100.0) | 17 (100.0) | 15 (88.2) | |
| High | 0 (0.0) | 0 (0.0) | 2 (11.8) | |
| Previous counseling provision, n (%) | | | | $\chi^2$ = 0.515; p = 0.773 |
| Yes | 6 (35.3) | 7 (41.2) | 5 (29.4) | |
| No | 11 (64.7) | 10 (58.8) | 12 (70.6) | |

**Note.** Mean ± SD: Mean ± standard deviation; $\chi^2$: Chi-square test; F: One-way analysis of variance.

According to the results of the mixed repeated-measures ANOVA, the time effect for participants' SABS scores was found to be statistically significant ($F = 113.31$, $p < 0.001$). However, the group effect ($F = 0.836$, $p = 0.440$) and the group × time interaction ($F = 0.304$,

p = 0.739) were not significant. There was no difference between the groups in terms of Satisfaction in learning, Self-confidence in learning, and Skill scores ($p > 0.05$) (Table 3).

**Table 3. Comparison of the Scale Scores of the Study Groups (n = 51)**

| Variables | Virtual Patient Group (n = 17) | Face-to-Face Peer Group (n = 17) | Virtual Peer Group (n = 17) | Test value / p value |
|---|---|---|---|---|
| | Mean ± SD (Median; Min–Max) | Mean ± SD (Median; Min–Max) | Mean ± SD (Median; Min–Max) | |
| SABS 1 | 30.71 ± 4.79 (32; 20–38) | 31.00 ± 5.89 (31; 20–43) | 33.24 ± 6.88 (32; 22–46) | $F^a$ = 0.929; p=0.402 |
| SABS 2 | 24.00 ± 5.30 (24; 12–33) | 23.35 ± 4.95 (22; 16–32) | 25.24 ± 6.22 (26; 13–39) | $F^a$ = 0.511; p=0.603 |
| Time[b] | | | | F = 113.310; p<0.001 |
| Group | | | | F = 0.836; p=0.440 |
| Group × time | | | | F = 0.304; p=0.739 |
| Satisfaction in learning | 4.88 ± 0.14 (5; 4.6–5) | 4.80 ± 0.32 (5; 4–5) | 4.80 ± 0.29 (5; 4–5) | $KW^c$ = 0.127; p=0.939 |
| Self-confidence in learning | 4.62 ± 0.35 (4.71; 4–5) | 4.61 ± 0.34 (4.71; 3.71–5) | 4.53 ± 0.36 (4.42; 3.71–5) | $F^a$ = 0.334; p=0.718 |
| Skill score | 24.59 ± 4.00 (25; 15–30) | 23.76 ± 2.63 (23; 20–28) | 23.88 ± 4.31 (24; 12–29) | $F^a$ = 0.243; p=0.785 |

**Note:** SD: Standard deviation [a] One Way ANOVA; [b] Mixed Repeated Measures ANOVA; [c]Kruskal Wallis test

## 4. DISCUSSION

In this study, it was determined that students' scores for negative attitudes and beliefs regarding the provision of sexual counseling decreased significantly in all three groups after structured sexual health education including simulation; however, no significant difference was found in terms of group and group–time interaction. This finding shows that virtual patient simulation, peer simulation in a virtual environment, and face-to-face peer simulation methods used in sexual counseling education during pregnancy had similar effects in reducing students' negative attitudes and beliefs regarding the provision of sexual counseling. Although these studies did not involve simulation-based interventions, previous research has shown that sexual education is effective in improving healthcare professionals' negative attitudes and beliefs. Tuğut and Gölbaşı (2017) reported that sexual health assessment education provided a positive change in nursing students' Sexual Attitudes and Beliefs Scale scores (Tugut & Golbasi, 2017). In a study evaluating sexual health education based on the PLISSIT model, it was reported that the education increased nursing students' knowledge, attitudes, and self-efficacy levels and was effective in reducing sexual myths (Gündüz & Demirci, 2026). Similarly, it was stated that a sexual health care education program increased psychiatric nurses' self-confidence and generally improved their sexual knowledge and attitudes (Lu et al., 2024). Another study found that peer education on sexual health assessment among nursing students had a positive effect on attitudes toward sexual health (Çulha & Afşin, 2023). Contrary to these findings, in a study conducted by Doğan et al. (2022) with the participation of nursing and midwifery students, it was stated that taking a sexual health course did not affect attitudes and beliefs regarding sexual health care; and it was recommended that interactive teaching methods such as role-playing, drama, and similar methods be integrated into education to make these courses more effective (Doğan et al., 2022). These findings suggest that theoretical instruction alone may not be sufficient to change attitudes and beliefs, and that students may need learning experiences that involve active participation and practice. In the present study, the fact that all three different simulation methods reduced negative attitudes and beliefs in providing sexual counseling indicates that interactive and practice-based educational approaches may make an important contribution in a sensitive area such as sexual counseling during pregnancy.

No significant difference was found between the groups in terms of sexual counseling skill scores. Although students in the virtual patient group interacted with a virtual patient model rather than a real individual, it is noteworthy that their counseling skill scores did not differ significantly from those of the peer simulation groups. This result suggests that virtual patient simulation supported by well-structured scenarios, synchronous responses, and verbal reactions may be a feasible alternative to peer simulation in areas requiring communication skills, such as sexual counseling during pregnancy. To our knowledge, this is the first study to evaluate the effect of a virtual patient simulation intervention in providing sexual counseling. In addition, although not randomized controlled trials, there are a limited number of studies on simulations such as peer/standardized patients in the context of sexual counseling. In a qualitative study conducted with gynecologic oncology nurses, nurses were asked to conduct an online interview focused on sexual health communication with a simulated participant. Following the simulation practice, participants reported positive developments in knowledge, beliefs, attitudes, communication skills, and nursing practices (Mrad et al., 2024). In another qualitative study, students assumed both the "healthcare professional" and "patient" roles through PLISSIT-based paired practices aimed at teaching sexual history-taking skills in an online environment. As a result of the study, themes such as becoming more comfortable using sexual language, using simpler expressions appropriate for the patient, increased self-confidence and sense of mastery, awareness of one's own body language and performance, structuring sexual history-taking, and integrating the PLISSIT model into practice were reported (Ross et al., 2021). In a study conducted with nursing students, students interviewed a patient whose sexual intimacy was affected due to chronic illness via videoconference; after the practice, a significant increase in students' communication self-confidence and a significant decrease in anxiety scores were found (Patrick & Butzlaff, 2021). Although on a different topic, only one study comparing virtual patient simulation and peer simulation was found. In this study, which compared virtual patient simulation and peer simulation in family planning counseling education, communication skill scores were reported to be higher in the virtual patient simulation group, whereas family planning counseling skill scores were higher in the peer simulation group (Şimşek Çetinkaya et al., 2024). In the present study, however, no significant difference was found between the groups in terms of sexual counseling skill scores. This difference may have resulted from differences in the subject areas, scenario content, assessment tools, and simulation environments of the studies. Nevertheless, both studies show that virtual patient and peer simulations are effective educational methods for developing students' counseling skills. In addition, the absence of a significant difference between the groups in this study suggests that the use of online peers or virtual patients may be considered as an educational alternative in situations where face-to-face practices are not always possible. In this study, it was determined that there was no statistically significant difference between the groups in terms of students' satisfaction in learning scores. The literature indicates that simulation-based learning provides a safe and controlled educational environment that supports students' satisfaction with their learning experience (Bdiri Gabbouj et al., 2024; Foronda et al., 2020; Wong & Wong, 2023). In nursing education, simulation supports students' satisfaction and self-confidence in learning by allowing them to experience clinical situations without risking patient safety, transfer their knowledge and skills into practice, and actively participate in the learning process (Bdiri Gabbouj et al., 2024; Wong & Wong, 2023). In addition, studies on virtual patient simulations also emphasize that this method may be acceptable to learners. Virtual patient simulations support satisfaction and learning by enabling students to encounter scenarios similar to real clinical situations, practice decision-making and communication skills in a safe environment, and experience the learning process in a structured manner (Mahou et al., 2023; Martini et al., 2019). Accordingly, the finding of the present study suggests that virtual patient simulation is a method that can support students' satisfaction with learning in sexual counseling education during pregnancy and may be considered as an alternative to peer simulations.

In this study, no significant difference was found between the groups in terms of students' self-confidence in learning scores. In the literature, there are studies reporting that nursing students' self-confidence in learning is high after simulation experiences (Alharbi & Alharbi, 2022; Moreno-Cámara et al., 2024; Toqan et al., 2023). In our study, the finding that the self-confidence scores of the virtual patient simulation group were similar to those of the peer simulation groups suggests that virtual patient simulation may be an acceptable and effective method at a level comparable to peer simulations in promoting students' self-confidence in learning. One study compared virtual patients with mannequin-based training in terms of student satisfaction and self-confidence in learning and showed no difference between the simulation methods (Haerling, 2018). In a systematic review, it was reported that virtual simulation increased students' self-confidence in the majority of studies; however, some studies comparing virtual simulation with traditional educational methods found no significant difference in terms of self-confidence (Foronda et al., 2020). In the study by Şimşek Çetinkaya et al., students also stated that both virtual patient simulation and peer simulation motivated learning, increased their self-confidence, and provided an enjoyable learning environment (Şimşek Çetinkaya et al., 2024). In the study by Karaduman and Başak comparing virtual patient simulation with human patient simulation, positive changes were observed in nursing anxiety and clinical decision-making self-confidence scores in all groups; virtual patient simulation was reported to yield superior results compared with the other methods in terms of nursing anxiety, clinical decision-making self-confidence, simulation-based learning, and performance scores (Karaduman & Basak, 2023). When these results are considered together, it can be stated that virtual patient simulation may be an acceptable and effective method at a level comparable to peer simulations in supporting students' self-confidence in learning.

**Conclusion**

This study showed that virtual patient simulation, peer simulation in a virtual environment, and face-to-face peer simulation methods used in sexual counseling education during pregnancy were effective in reducing students' negative attitudes and beliefs regarding providing sexual counseling. However, no significant difference was found between the methods in terms of attitudes and beliefs regarding sexual counseling, sexual counseling skills, satisfaction in learning, and self-confidence in learning. These findings suggest that different simulation approaches can provide similar educational outcomes and that virtual patient simulation may be a feasible alternative to traditional peer simulations. In addition, the absence of significant differences between the groups indicates that online peer simulation and virtual patient simulation can be used as effective educational alternatives in situations where face-to-face education is not possible or is limited. Accordingly, it is recommended that the use of interactive methods such as virtual patient simulation and peer simulation be expanded in nursing education programs for teaching topics that require communication skills, such as sexual counseling during pregnancy.

**References**

Açıkgöz, G., & Baykal, Ü. (2023). Legal regulations supporting the professional roles and autonomy of nurses. *Istanbul Kent University Journal of Health Sciences*, *2*(1), 29–34.

Alharbi, K., & Alharbi, M. F. (2022). Nursing students' satisfaction and self-confidence levels after their simulation experience. *SAGE Open Nursing*, *8*, 23779608221139080. https://doi.org/https://doi.org/10.1177/23779608221139080

Altınayak, S. Ö., Apay, S. E., & Vermeulen, J. (2020). The role of midwifery associations in the professional development of midwifery. *European Journal of Midwifery*, *4*, 27.

Annon, J. S. (1976). The PLISSIT model: a proposed conceptual scheme for the behavioral treatment of sexual problems. *Journal of sex education and therapy*, *2*(1), 1–15.

Ayhan, H., Iyigun, E., Tastan, S., & Coskun, H. (2010). Turkish version of the reliability and validity study of the sexual attitudes and belief survey. *Sexuality and Disability*, *28*(4), 287–296.

Bdiri Gabbouj, S., Zedini, C., & Naija, W. (2024). Nursing students' satisfaction and self-confidence with simulation-based learning and its associations with simulation design characteristics and educational practices. *Advances in Medical Education and Practice*, 1093–1102. https://doi.org/https://doi.org/10.2147/AMEP.S477309

Bensing, J. M., Visser, A., & Saan, H. (2001). Patient education in the Netherlands. *Patient education and Counseling*, *44*(1), 15–22. https://doi.org/https://doi.org/10.1016/S0738-3991(01)00097-0

Bogaert, E., & Roels, R. (2025). Sexual health in patient care: shortcomings in medical training and experienced barriers in sexual history taking. *BMC Medical Education*, *25*(1), 338.

Borms, R., & Vermeire, K. (2020). Spreken is goud: Seksuele gezondheid bespreekbaar maken met de Onder 4 ogen methode. Ontwikkeling en implementatie bij huisartsen in Vlaanderen. *Tijdschrift voor Seksuologie*, *44*(3), 155–161.

Bozkurt Karalı, M. N., & Özkan, Y. (2020). Counselor Self Efficacy Scale: The Adaptation Study into Turkish. *Turkiye Klinikleri Journal of Health Sciences*, *5*(3), 616–624.

Bracq, M.-S., Michinov, E., & Jannin, P. (2019). Virtual reality simulation in nontechnical skills training for healthcare professionals: a systematic review. *Simulation in Healthcare*, *14*(3), 188–194.

Çulha, Y., & Afşin, F. (2023). The Effect of Peer Education about Assessment of Sexual Health in Nursing Students on their Attitudes towards Sexual Health. *University of Health Sciences Journal of Nursing*, *5*(1), 23–28. https://doi.org/https://doi.org/10.48071/sbuhemsirelik.1230476

Doğan, N., Fışkın, G., & Yüceler Kaçmaz, H. (2022). Beliefs and attitudes regarding sexual health care of students who take and
didn't take sexual health lessons. *Androl Bul*, *24*(1).

Fathalian, M., Lotfi, R., Faramarzi, M., & Qorbani, M. (2022). The effect of virtual cognitive-behavioral sexual counseling on sexual function and sexual intimacy in pregnant women: a randomized controlled clinical trial. *BMC Pregnancy and Childbirth*, *22*(1), 616. https://doi.org/https://doi.org/10.1186/s12884-022-04932-4

Fennell, R., & Grant, B. (2019). Discussing sexuality in health care: A systematic review. *Journal of Clinical Nursing*, *28*(17-18), 3065–3076.

Ferrara, E., Pugnaire, M. P., Jonassen, J. A., O'Dell, K., Clay, M., Hatem, D., & Carlin, M. (2003). Sexual health innovations in undergraduate medical education. *International* *journal of impotence research*, *15*(5), S46–S50. https://doi.org/https://doi.org/10.1038/sj.ijir.3901072

Foronda, C. L., Fernandez-Burgos, M., Nadeau, C., Kelley, C. N., & Henry, M. N. (2020). Virtual simulation in nursing education: a systematic review spanning 1996 to 2018. *Simulation in Healthcare*, *15*(1), 46–54.

González, B., Lupón, J., Herreros, J., Urrutia, A., Altimir, S., Coll, R., Prats, M., & Valle, V. (2005). Patient's education by nurse: what we really do achieve? *European Journal of Cardiovascular Nursing*, *4*(2), 107–111.

Gündüz, C. S., & Demirci, N. (2026). The Effects of Distance Sexual Health Education Based on the PLISSIT Model on Knowledge, Attitude, And Self-Efficacy in Nursing Students: A Randomized Controlled Trial. *International Journal of Sexual Health*, *38*(1), 239–247.

Haerling, K. A. (2018). Cost-utility analysis of virtual and mannequin-based simulation. *Simulation in Healthcare*, *13*(1), 33–40.

Henning, J. M., Weidner, T. G., & Marty, M. C. (2008). Peer assisted learning in clinical education: Literature review. *Athletic training education journal*, *3*(3), 84–90. https://doi.org/https://doi.org/10.4085/1947-380x-3.3.84

Jeffries, P. R., & Rizzolo, M. A. (2006). *Designing and implementing models for the innovative use of using simulation to teach nursing care of Ill adults and children: A national, multi-site, multi-method study*. National League for Nursing.

Karaduman, G. S., & Basak, T. (2023). Is virtual patient simulation superior to human patient simulation: a randomized controlled study. *CIN: Computers, Informatics, Nursing*, *41*(6), 467–476.

Kelder, I., Sneijder, P., Klarenbeek, A., & Laan, E. (2022). Communication practices in conversations about sexual health in medical healthcare settings: A systematic review. *Patient education and Counseling*, *105*(4), 858–868.

Kienle, R., Freytag, J., Lück, S., Eberz, P., Langenbeck, S., Sehy, V., & Hitzblech, T. (2021). Communication skills training in undergraduate medical education at Charité–Universitätsmedizin Berlin. *GMS Journal for Medical Education*, *38*(3), Doc56.

Kong, S. K. F., Wu, L. H., & Loke, A. Y. (2009). Nursing students' knowledge, attitude and readiness to work for clients with sexual health concerns. *Journal of Clinical Nursing*, *18*(16), 2372–2382.

Kononowicz, A. A., Woodham, L. A., Edelbring, S., Stathakarou, N., Davies, D., Saxena, N., Car, L. T., Carlstedt-Duke, J., Car, J., & Zary, N. (2019). Virtual patient simulations in health professions education: systematic review and meta-analysis by the digital health education collaboration. *Journal of medical Internet research*, *21*(7), e14676.

Lane, C., & Rollnick, S. (2007). The use of simulated patients and role-play in communication skills training: a review of the literature to August 2005. *Patient education and Counseling*, *67*(1-2), 13–20.

Lu, M.-J., Li, J.-B., Wu, C.-Y., Huong, P. T. T., Hsu, P.-C., & Chang, C.-R. (2024). Effectiveness of a sexual health care training to enhance psychiatric nurses' knowledge, attitude, and self-efficacy: a quasi-experimental study in southern Taiwan. *Journal of the American Psychiatric Nurses Association*, *30*(1), 17–29.

Magill, M., Mastroleo, N. R., & Martino, S. (2022). Technology-based methods for training counseling skills in behavioral health: A scoping review. *Journal of technology in behavioral science*, *7*(3), 325–336.

Mahou, F., Elamari, S., Sulaiman, A. A., Bouaddi, O., Changuiti, O., Mouhaoui, M., & Khattabi, A. (2023). Teaching nursing management of diabetic ketoacidosis: a description of the development of a virtual patient simulation. *Advances in Simulation*, *8*(1), 2. https://doi.org/https://doi.org/10.1186/s41077-022-00241-0

Manninen, S.-M., Kero, K., Riskumäki, M., Vahlberg, T., & Polo-Kantola, P. (2022). Medical and midwifery students need increased sexual medicine education to overcome barriers hindering bringing up sexual health issues–A national study of final-year medical and midwifery students in Finland. *European Journal of Obstetrics & Gynecology and Reproductive Biology*, *279*, 112–117.

Manninen, S.-M., Polo-Kantola, P., Riskumäki, M., Vahlberg, T., & Kero, K. (2024). The knowledge of and educational interest in sexual medicine among Finnish medical and midwifery students: A web-based study. *European Journal of Midwifery*, *8*, 10.18332/ejm/186401.

Martini, N., Farmer, K., Patil, S., Tan, G., Wang, C., Wong, L., & Webster, C. S. (2019). Designing and evaluating a virtual patient simulation—the journey from uniprofessional to interprofessional learning. *Information*, *10*(1), 28. https://doi.org/https://doi.org/10.3390/info10010028

Merhavy, Z., Varkey, T., Kotyk, T., & Zeitler, C. (2023). Sexual health preparedness among medical students. *Медичні перспективи= Medicni perspektivi (Medical perspectives)*(4), 129–140. https://doi.org/https://doi.org/10.26641/2307-0404.2023.4.294193

Mick, J., Hughes, M., Cohen, M. Z., & Brant, J. M. (2004). Using the BETTER Model to assess sexuality. *Clinical journal of oncology nursing*, *8*(1).

Moreno-Cámara, S., da-Silva-Domingues, H., Parra-Anguita, L., & Gutiérrez-Sánchez, B. (2024). Evaluating satisfaction and self-confidence among nursing students in clinical simulation learning. *Nursing Reports*, *14*(2), 1037–1048. https://doi.org/https://doi.org/10.3390/nursrep14020078

Mrad, H., Chouinard, A., Pichette, R., Piché, L., & Bilodeau, K. (2024). Feasibility and impact of an online simulation focusing on nursing communication about sexual health in gynecologic oncology. *Journal of Cancer Education*, *39*(1), 3–11.

Patrick, S. R., & Butzlaff, A. (2021). By utilizing technology can nursing students gain more confidence and decrease anxiety when communicating with chronically ill patients about their sexual relationship? *Nurse education today*, *107*, 105084. https://doi.org/https://doi.org/10.1016/j.nedt.2021.105084

Peddle, M., Bearman, M., & Nestel, D. (2016). Virtual patients and nontechnical skills in undergraduate health professional education: an integrative review. *Clinical Simulation in Nursing*, *12*(9), 400–410.

Pouresmail, Z., Nabavi, F. H., & Zare, N. V. (2023). Outcomes of patient education in nurse-led clinics: a systematic review. *Journal of Caring Sciences*, *12*(3), 188.

Reynolds, K. E., & Magnan, M. A. (2005). Nursing attitudes and beliefs toward human sexuality: Collaborative research promoting evidence-based practice. *Clinical Nurse Specialist*, *19*(5), 255–259.

Ross, M. W., Newstrom, N., & Coleman, E. (2021). Teaching sexual history taking in health care using online technology: a PLISSIT-plus zoom approach during the coronavirus disease 2019 shutdown. *Sexual Medicine*, *9*(1), 100290–100290.

Sezer, B., Sezer, T. A., Teker, G. T., & Elcin, M. (2023). Developing a virtual patient: design, usability, and learning effect in communication skills training. *BMC Medical Education*, *23*(1), 891.

Shahbazi, Z., Farshbaf, K. A., Sattarzadeh, N., & Kamalifard, M. (2019). The effect of sexual counseling based on PLISSIT model on sexual function of pregnant women: a randomized controlled clinical trial.

Shatto, B., Meyer, G., Krieger, M., Kreienkamp, M. J., Kendall, A., & Breitbach, N. (2022). Educational interventions to improve graduating nursing students' practice readiness: a systematic review. *Nurse educator*, *47*(2), E24.

Şenyuva, E., & Akince, E. K. (2020). Is peer education an effective method of strengthening nursing education? Akran eğitimi hemşirelik eğitimini güçlendirmede etkili bir yöntem midir? *Journal of Human Sciences*, *17*(1), 92–103.

Şimşek Çetinkaya, Ş., Gümüş Çalış, G., Kıbrıs, Ş., & Topal, M. (2024). Effectiveness of virtual patient simulation versus peer simulation in family planning training in midwifery students: a comparative educational intervention. *Interactive Learning Environments*, *32*(3), 942–951.

Taylor, B., & Davis, S. (2006). Using the extended PLISSIT model to address sexual healthcare needs. *Nurs Stand.*, *21*(11). https://doi.org/10.7748/ns2006.11.21.11.35.c6382

Toqan, D., Ayed, A., Khalaf, I. A., & Alsadi, M. (2023). Effect of high-fidelity simulation on self-satisfaction and self-confidence among nursing students. *SAGE Open Nursing*, *9*, 23779608231194403. https://doi.org/https://doi.org/10.1177/23779608231194403

Tugut, N., & Golbasi, Z. (2017). Sexuality Assessment Knowledge, Attitude, and Skill of Nursing Students: An Experimental Study with Control Group. *International Journal of Nursing Knowledge*, *28*(3).

Unver, V., Basak, T., Watts, P., Gaioso, V., Moss, J., Tastan, S., Iyigun, E., & Tosun, N. (2017). The reliability and validity of three questionnaires: the student satisfaction and self-confidence in learning scale, simulation design scale, and educational practices questionnaire. *Contemporary nurse*, *53*(1), 60–74.

Verrastro, V., Saladino, V., Petruccelli, F., & Eleuteri, S. (2020). Medical and health care professionals' sexuality education: state of the art and recommendations. *International journal of environmental research and public health*, *17*(7), 2186.

Wong, F. M., & Wong, D. C. (2023). A modified guideline for High-Fidelity patient simulation to improve student satisfaction and Self-Confidence in learning: A mixed study. *Nursing Reports*, *13*(3), 1030–1039. https://doi.org/https://doi.org/10.3390/nursrep13030090

Yang, J., & Kim, S. (2022). An online communication skills training program for nursing students: A quasi-experimental study. *Plos one*, *17*(5), e0268016.

Yılancıoğlu, H. Y., & Bildik, T. (2022). Applications of Virtual Patient. *Turkiye Klinikleri Child Psychiatry-Special Topics*, *8*(1), 136–140.

Yoo, M. S., & Chae, S.-M. (2011). Effects of peer review on communication skills and learning motivation among nursing students. *Journal of Nursing Education*, *50*(4), 230–233. https://doi.org/https://doi.org/10.3928/01484834-20110131-03

Ziaei, T., Kharaghani, R., Haseli, A., & Ahmadnia, E. (2022). Comparing the effect of extended PLISSIT model and group counseling on sexual function and satisfaction of pregnant women: A randomized clinical trial. *Journal of Caring Sciences*, *11*(1), 7. https://doi.org/https://doi.org/10.34172/jcs.2022.06

Zolkefli, Y., & Mahmud, M. H. (2023). Ethical Responsibilities in Patient Education. *INTERNATIONAL JOURNAL OF CARE SCHOLARS*, *6*(2), 69–70. https://doi.org/https://doi.org/10.31436/ijcs.v6i2.312